# Empowering Learner-Centered Instruction: Integrating ChatGPT Python API and Tinker Learning for Enhanced Creativity and Problem-Solving Skills

Yun-Cheng Tsai[1]

National Taiwan Normal University, Taipei 10610, Taiwan, R.O.C.

**Abstract.** The ChatGPT Python API plays a crucial role in promoting Learner-Centered Instruction (LCI) and aligns with the principles of Tinker Learning, allowing students to discover their learning strategies. LCI emphasizes the importance of active, hands-on learning experiences and encourages students to take responsibility for their learning journey. By integrating the ChatGPT Python API into the educational process, students can explore various resources, generate new ideas, and create content in a more personalized manner. This innovative approach enables students to engage with the learning material deeper, fostering a sense of ownership and motivation. As they work through the Creative Learning Spiral, students develop essential skills such as critical thinking, problem-solving, and creativity. The ChatGPT Python API is a valuable tool for students to explore different solutions, evaluate alternatives, and make informed decisions, all while encouraging self-directed learning. In Tinker Learning environments, the integration of ChatGPT Python API empowers students to experiment and iterate, allowing them to find the most effective learning strategies that cater to their individual needs and preferences. This personalized approach helps students to become more confident in their abilities, leading to tremendous academic success and long-term skill development. By leveraging the capabilities of the ChatGPT Python API, educational institutions can create a more engaging, supportive, and dynamic learning environment. This approach aligns with the principles of Learner-Centered Instruction and Tinker Learning, promoting a culture of curiosity, exploration, and creativity among students while preparing them for the challenges of the fast-paced, ever-changing world.

## 1 Introduction

Learner-Centered Instruction (LCI) fosters active learning experiences and empowers students to take charge of their educational journey [21]. This instructional approach emphasizes hands-on exploration, problem-solving, and collaboration as essential for knowledge construction [15]. LCI cultivates a culture of curiosity, exploration, and creativity, equipping students to face the rapidly evolving, dynamic world [6]. The Cone of Learning model posits that the most effective learning occurs through firsthand experiences, supplemented by practicing with the material, hearing about it, and lastly, reading about it [5].

The Tinker Learning approach advocates for learners to build their knowledge through hands-on exploration and discovery [18]. Grounded in the notion that students learn optimally by experimenting with and manipulating learning materials instead of receiving direct instructions [12], this approach closely aligns with the Cone of Learning, underscoring firsthand experience and practice as pivotal to effective learning [8]. The Creative Learning Spiral, a five-step process, guides learners through creative problem-solving [17], fostering creativity and promoting out-of-the-box thinking during challenges [19].

ChatGPT is a natural language processing technology developed by OpenAI that aims to generate fluent and coherent text. The API for this technology became available to the public in June 2020, allowing developers and researchers to harness the power of ChatGPT easily. ChatGPT is a commercial product released by OpenAI. Using the ChatGPT API is subject to their terms of service and policies [16]. We have extensively utilized the ChatGPT API in my classroom for cross-disciplinary text mining and programming teaching. This approach has proved effective in increasing students' sense of achievement and breaking through the sample size limitations in qualitative analysis. By harnessing the power of ChatGPT, students can engage in deep thinking through the practical application of programming skills, leading to a more profound understanding of the subject matter. Moreover, the ChatGPT API has enabled us to explore new avenues in text analysis and language processing. It has provided students with the opportunity to learn cutting-edge technologies and techniques. Through this teaching methodology, students have developed a strong foundation in programming, critical thinking, and problem-solving skills, equipping them with valuable skills essential for success in today's rapidly evolving technological landscape. As such, using the ChatGPT API in the classroom is a promising pedagogical approach that can lead to significant educational benefits for educators and students.

Incorporating resources such as the ChatGPT Python API, this paper delineates a process involving imagination, creation, play, sharing, and feedback reception. The strategy aligns with the Tinker Learning approach and the Cone of Learning, emphasizing the significance of hands-on exploration, experimentation, and trial-and-error in learning [4]. LCI is compatible with educational models like the Cone of Learning and Tinker Learning, which accentuate firsthand experiences and practical application in learning processes [8]. These three models underscore the importance of active engagement with learning through experimentation and hands-on activities. They concur that students learn most effectively when actively involved in and encouraged to explore and construct new knowledge or skills [14]. Each model offers distinct perspectives on learning, focusing on retention [7], hands-on activities [18], and creativity cultivation [17], respectively. By understanding LCI's advantages and alignment with models such as the Cone of Learning and Tinker Learning, educators can make well-informed decisions regarding the most efficacious classroom management strategies for their students' success [11].

The structure of this paper is as follows: in Section 2, We review previous research on computer science education and learner sourcing. In Section 3, We describe our research methods and research questions. The results are presented in Section 4 and discussed in Section 5. Finally, we conclude the article in Section 6.

## 2 Literature Review

### 2.1 Learner-Centered Instruction (LCI)

In the recent ten years, Learner-Centered Instruction (LCI) has gained increasing prominence in education [22]. The World Economic Forum's "Education 4.0" initiative, released in May 2022, emphasizes the importance of a learner-centered teaching model that utilizes technology and innovation to equip learners with diverse skills to navigate the challenges of the fourth industrial revolution [20]. The Cone of Learning model illustrates the learning levels based on retention and difficulty. It suggests that learners retain more information when actively learning instead of passively receiving information [9].

Therefore, learner-centered instruction, which puts the learner at the center of the learning process and allows them to be actively engaged, can effectively combine this model with a new learning strategy. This approach encourages learners to take ownership of their learning and develop higher-order thinking skills, such as analysis, synthesis, evaluation, and application. By incorporating problem-based, collaborative, and inquiry-based learning into the curriculum, learners can engage in meaningful and authentic learning experiences that can result in a deeper understanding and retention of the material [1][20]

### 2.2 Tinker Learning

Research has shown that learner-centered approaches and related teaching strategies, such as the Tinker Learning method proposed by Seymour Papert, a professor at MIT, are effective in improving learning outcomes [13][3].

Tinker Learning is a learner-centered instruction strategy incorporating the Cone of Learning concept to create a new approach to learning [10]. The Cone of Learning theory suggests that people retain more information when actively involved in the learning process rather than passively receiving information. By combining these two concepts, Tinker Learning allows learners to actively explore and experiment with new ideas and concepts, leading to a deeper understanding and retention of the material. Through this process, learners can construct their knowledge and develop their own "learning how to learn" and 4C (creativity, critical thinking, communication, and collaboration) skills [2].

### 2.3 Creative Learning Spiral

The Tinker Learning approach, also known as the Creative Learning Spiral, involves a process of playful exploration, creation, and iteration in which students develop their knowledge through the repeated cycles of imagining, creating, executing, and completing projects, as well as sharing and receiving feedback [17]. This process is similar to children's play, in which they experiment with different combinations and configurations, seeking new ideas and expressing their creativity through exploration and experimentation.

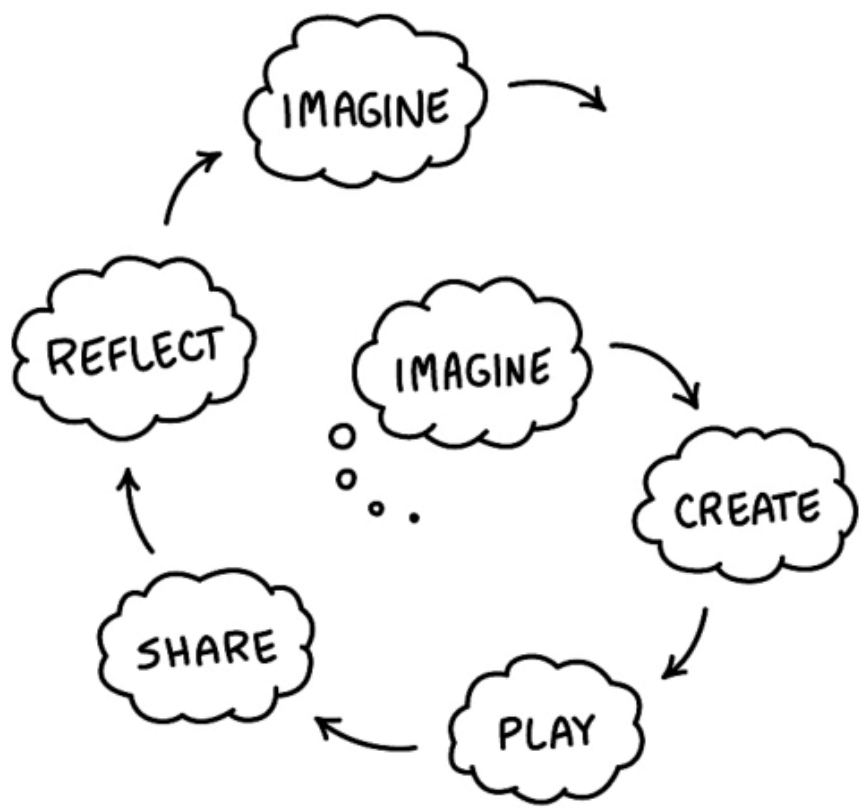

**Fig. 1.** Creative Learning Spiral by Mitchel Resnick

A creative Learning Spiral is a teaching strategy that guides students through the process of creative thinking, from idea generation to implementation. Fig. 1 involves five steps as follows:

1. IMAGINE: trying out new ideas.
2. CREATE: exploring different solution paths.
3. PLAY: making adjustments.
4. SHARE: imagining new possibilities.
5. REFLECT: creatively expressing ideas.

Students can build up their knowledge base in a spiral fashion by repeatedly attempting to develop new ideas, create, explore through play, and share and receive feedback. This approach helps students understand how creative ideas are generated and develops their skills as creative thinkers.

## 3  Methods

The paper's approach to teaching students the skills of using ChatGPT Python API is based on the principles of Tinker Learning, an approach developed by MIT Professor Seymour Papert. The critical principle of Tinker Learning is that it emphasizes providing learners with "sufficient learning environments" in which they can construct their knowledge. The role of the teacher shifts from simply lecturing and demonstrating fixed examples to serving as a coach or mentor, and the classroom becomes more like a "swimming pool or sports field." The following process is an example of Methods for Analyzing Highly Cited Blockchain in Education Related Papers from 2019-2023 Using ChatGPT and LDA.

1. Data Collection: We collected highly cited Blockchain in Education-related papers published from 2019-2023 through academic search engines such as Google Scholar, Microsoft Academic, and Semantic Scholar.
2. ChatGPT Python API: We utilized ChatGPT Python API, a language generation model, to generate summaries for each paper. These summaries provided a brief understanding of the content of each paper.
3. LDA: We applied LDA, a topic modeling algorithm, to identify the main topics discussed in the papers. We set the number of issues and ran the model to obtain each topic's main word list and corresponding weights.
4. Analysis: The Latent Dirichelet Allocation (LDA) results were analyzed to identify the topics discussed in the highly cited Blockchain in Education-related papers. We also cross-checked the paper summaries to confirm the issues.
5. Conclusion: We used ChatGPT Python API and LDA to analyze highly cited Blockchain in Education-related papers from 2019-2023. Our analysis identified the main topics discussed in these papers, which will help to inform future research in this area.

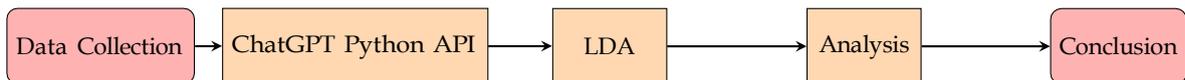

The length of the arrows in our teaching strategy represents the time it takes for both the teacher and student to progress to the next stage. With the help of ChatGPT, we aim to reduce the time spent on pre-processing text during qualitative analysis, allowing for more discussion and deep thinking after visualizing the data. We believe this approach can foster autonomous learning and analytical thinking skills among students, enabling them to take ownership of their learning experience. By utilizing the ChatGPT Python API, educators can empower students to delve deeper into their research topics and reach their full potential. This, combined with Tinker Learning's learner-centered teaching strategies, can help students acquire the critical "how to learn" skills necessary for success in a rapidly evolving technological landscape. To implement this approach in our classroom, we use live coding and debugging techniques to help students build their knowledge step by step as they practice writing programming code to solve real-world problems. We also use the GitHub version control platform to track students' code submissions and changes, allowing them to identify problems, design solutions, plan actions, collect data, make decisions, and present their work. This way, Tinker Learning transforms the traditional teacher-centered classroom into a more learner-centered one.

We found that the programming languages students learn in school may differ from those they encounter in the workforce. For example, in 2000, university students were mainly learning C language, as Python did not yet exist. However, through self-learning, we have become proficient in Python and can teach students how to apply it to problems of interest to us. This is because we have acquired the "how to learn" skills through our C language experience. We have been thinking for seven years about incorporating the skills of "how to learn" into a teaching strategy to help students "learn how to learn." It was not until we read Lifelong Kindergarten: Cultivating Creativity through Projects, Passion, Peers, and Play by Mitchel Resnick that we realized the core spirit of Tinker Learning's teaching strategies is that good education is not just about how well teachers can teach, but about providing learners with a "sufficient learning environment" to construct their own body of knowledge. By drawing an analogy between "learners constructing their own body of knowledge" and "how to learn" skills, we have found that Tinker Learning may be the solution we have been seeking. We are now interested in exploring how we can use Tinker Learning's teaching strategies to enable students to acquire the skills of "how to learn."

Fig. 2 shows that we use Python API for ChatGPT, which offers a solution to the character limit present in some user interface windows, allowing for the analysis of large amounts of text through loops and def functions. This feature has the potential to save time previously spent on the tedious task of preprocessing text. By freeing up this time, students can focus on the more critical functions of analysis, discussion, and data visualization. This approach encourages autonomous learning and fosters analytical thinking skills, empowering students to take ownership of their learning experience. By utilizing the ChatGPT API, educators can give students the tools to delve deeper into their research topics and reach their full potential.

### 3.1    Learning How to Learn

The goal of this teaching is "learning how to learn" and the 4Cs: critical thinking, communication, collaboration, and creativity. No one can honestly know how the future will change, and any assumptions

```
PUS database on the topic "Blockchain Technology in Education". The following research questions guided this systematic literature review (SLR: How blockchain technology has been
de.ned in educational settings? How were the technology examined (i.e. the methodology)? What were the results of using this technology in an education system? \nFindings - The s
tudy identi.es the bene.ts,' ' barriers and present application of blockchain technology in education. The analysis shows that blockchain technology in education is still a young
 discipline',' ' but has a lot of potential to bene.ts the educational sector at large. \nPractical implications - This research provides a groundwork for education institutions',
' the policymakers and researchers to explore other areas where blockchain technology can be implemented',' ' though this research has also suggested some prospective uses of block
chain technology in different functions of an education system',' ' more application can be brought into the education system to exploit the potential of blockchain technology. \n
 Originality/value - The paper discusses the application of blockchain technology in education with the help of bibliometric analysis. This is one of the .rst known studies to revi
ew the blockchain technology by identifying its bene.ts,' ' barriers',' ' present blockchain technology application. Based on the analysis',' ' future application areas are also ide
nti.ed. \nKeywords Systematic review',' ' Education',' ' Blockchain technology',' ' Blockchain application \nPaper type Literature review \n1. Introduction \nBlockchain was .rst
 used to a peer-to-peer ledger for record-keeping of the transactions of Bitcoin cryptocurrency. A blockchain transaction in the public ledger contains a veri.able record and once
  the information entered',' ' it cannot be altered or erased in the future. The \nBlockchain in education management \nReceived 4 July 2020 Revised 19 September 2020 Accepted 10 O
ctober 2020 \n\nInteractive Technology and Smart Education Vol. 18 No. 1',' ' 2021 pp. 1-17 © Emerald Publishing Limited 1741-5659 DOI 10.1108/ITSE-07-2020-0102 \n\Blockchain
tech"
```

```python
In [12]: def chatgptfn(sub_list):
             result = ''
             response = openai.ChatCompletion.create(
                 model="gpt-3.5-turbo",
                 messages=[
                     {"role": "system", "content": "You are an assistant"},
                     {"role": "user", "content": f"{sub_list} :give me a summary"}
                 ]
             )
             for choice in response.choices:
                 result += choice.message.content
             return result
```

```python
In [13]: data[0] = chatgptfn(data[0])
```

```python
In [14]: data[0]
```

```
Out[14]: 'This paper provides a literature review on the application of blockchain technology in education. The study examines the benefits and barriers associated with blockchain technolo
gy, as well as its current applications in the educational sector. Additionally, the paper identifies potential future areas in which blockchain technology can be implemented. The
research is based on bibliometric analysis and provides a foundation for educators, policymakers, and researchers to explore the potential of blockchain technology in education.'
```

```python
In [15]: data[1] = chatgptfn(data[1])
```

```python
In [9]: data[2] = chatgptfn(data[2])
        data[3] = chatgptfn(data[3])
```

```python
In [18]: for i in range(0,5):
             data[i] = chatgptfn(data[i])
```

```python
In [19]: import gensim
         from gensim import corpora
         from pprint import pprint
```

```python
In [20]: # 創建詞袋
         texts = [[word for word in document.lower().split()] for document in data]
         dictionary = corpora.Dictionary(texts)
         corpus = [dictionary.doc2bow(text) for text in texts]
```

```python
In [21]: # 訓練 LDA 模型
         lda_model = gensim.models.ldamodel.LdaModel(corpus=corpus, id2word=dictionary,
                                                     num_topics=3, random_state=100, update_every=1,
                                                     chunksize=100, passes=10, alpha='auto', per_word_topics=True)
```

**Fig. 2.** The example of Methods for Analyzing Highly Cited Blockchain in Education Related Papers from 2019-2023 Using ChatGPT and LDA.

may be far from the actual end. Therefore, what schools should give students now should be "learning how to learn" and the 4Cs: critical thinking, communication, collaboration, and creativity. More broadly, the curriculum should emphasize general information skills that can be integrated into daily life and, most importantly, the ability to adapt, learn new things, maintain mental balance in unfamiliar situations, and find appropriate solutions to problems. In such a world, "Less is more." Teachers do not need to teach students more information. Students must understand data, judge what is essential, and combine bits of information to form a holistic worldview. All technical hands-on courses using programming languages as the development tool are the best way to approach such a world and achieve the 4Cs. Like the brain bookshelf in the picture below, after the teacher gives students the minimum essential tools, the rest of the knowledge and ability development is for students to build their learning through hands-on work.

Good education is not just about making teachers teach well but providing a "sufficient learning environment" for learners to construct their knowledge. We found that Tinker Learning was the answer we had been looking for. We want to find out how we acquired the skills of "how to learn" through this project and to use the teaching strategies of Tinker Learning to enable students to develop the skills of "how to learn."

### 3.2 Tinker Learning, Live Coding, and Live debugging

The goal of teaching is "learning how to learn" and the 4Cs. Because no one can honestly know how the future will change and any assumptions may be far from reality, schools should focus on teaching students "how to learn" and the 4Cs: critical thinking, communication, collaboration, and creativity. The curriculum should emphasize general information skills that can be integrated into daily life and, most importantly, the ability to adapt, learn new things, maintain mental balance in unfamiliar situations, and find appropriate solutions to problems. In such a world, "less is more." Instead of overwhelming students with information, it is more critical for them to understand data, judge what information is essential, and combine bits of information to form a holistic view of the world. Technical, hands-on courses using programming languages as development tools can provide an excellent approach to achieving these goals.

Tinker Learning, Live Coding, and Live debugging are teaching methods in this course. Tinker Learning involves a process of playful exploration, creation, and iteration. It is similar to children's play, where students experiment with different combinations and configurations, seeking new ideas and expressing their creativity through exploration and experimentation. This process helps students develop their knowledge through repeated cycles of imagining, creating, executing, and completing projects as sharing and receiving feedback. Live Coding and Live Debug involve students writing code to solve real-world problems, practicing identifying problems, designing projects, planning actions, collecting data, solving problems, and making decisions. These methods transform the traditional teacher-centered classroom into a more learner-centered environment, creating a "co-learning and co-creation" atmosphere at the school where the role of the teacher shifts from simply presenting fixed examples to serving as a coach. Using GitHub, teachers can help students track code submissions and changes, fostering the development of cross-disciplinary professionals with practical implementation skills. Tinker Learning promotes active, hands-on learning and encourages students to take ownership of their learning process through live coding, debugging, and tracking code submissions and changes via the GitHub platform.

### 3.3 Implementation of the Study

A program to analyze the records of all students on GitHub following the research framework. The teacher should guide the students to find a problem in their daily lives that interests them and discuss how to use the tools learned in the week to solve the problem. The teacher should demonstrate how they used specific techniques and tools to solve the problem and how these techniques can be extended. Through the Tinker Learning teaching strategy, students can try new ideas, explore different paths to solve problems, make adjustments, imagine new possibilities, and creatively express their views. The Creative Learning Spiral process allows students to understand how creativity develops from idea to implementation and become creative thinking practitioners. By repeating the process of trying to imagine, create, execute, and complete creations through play, sharing, and receiving feedback, students' knowledge is built up step by step like a spiral.

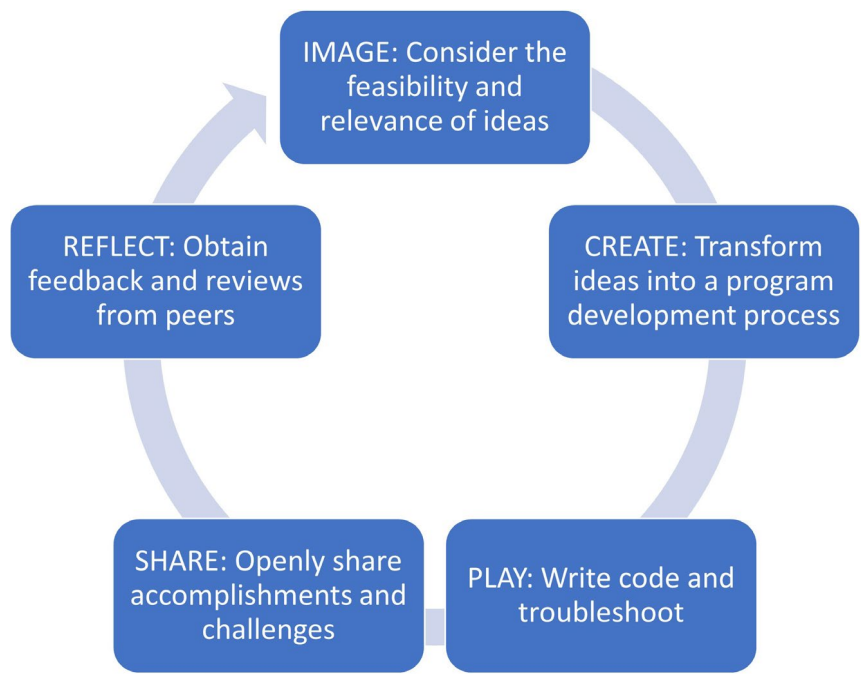

**Fig. 3.** Our Creative Learning Spiral for teaching students the skills of ChatGPT Python API.

Fig. 3 shows our Creative Learning Spiral for teaching the skills of ChatGPT Python API to students, incorporates the following five steps:

1. IMAGE: The ChatGPT Python API can assist students in exploring and refining their ideas by providing them with relevant information and insights based on their input.
2. CREATE: The ChatGPT Python API can be used as a tool to help students develop their projects by generating text and language models based on their requirements.
3. PLAY: Students can use the ChatGPT Python API to analyze and understand large amounts of text data, identify patterns and relationships, and draw insights from their findings. By working with the API, students can develop their analytical skills and gain a deeper understanding of the subject matter.
4. SHARE: GitHub can serve as a powerful platform for students to work together on coding projects, share code snippets, and provide feedback to one another. By incorporating the ChatGPT Python API in their projects, students can collaborate on analyzing large amounts of text data and gain insights from their findings.
5. REFLECT: By receiving feedback and reviews from their peers and instructors, students can reflect on what they have learned and how they can improve their skills in the future. The ChatGPT Python API can provide them with a valuable tool for identifying improvement areas and refining their problem-solving approach.

This approach emphasizes active engagement and experimentation as key components of effective learning and uses the Creative Learning Spiral process to guide students through creatively solving problems.

## 4 Results

This section will explore how students can demonstrate their proficiency and practical implementation skills by creating a portfolio of finished projects on their GitHub accounts. By analyzing the progress of their projects over time, we can track their growth and learn throughout the course. To achieve this, we will obtain all students' GitHub accounts, and using action research and case study methods, we will observe their weekly learning progress.

### 4.1 Participants' GitHub

Our approach involves comparing and analyzing changes in the students' code from the initial blank framework to the progress made during their first assignments and the final completed content. We can infer how they construct their knowledge by repeatedly verifying the students' assignments. The GitHub link containing the records of all 44 out of 45 students who made progress in the first three assignments through the tutorials is available at https://reurl.cc/7jzggN. It is evident that only two students did not receive full marks in the second assignment, and six students did not achieve full effects in the first assignment. Fig. 4 shows all participants' performance in assignments, which are available on GitHub sub-sheets, including HW1-HW3, HW4-HW5, and the final project. These assignments aim to confirm students' skills in data visualization, integrating programmatic skills, and exploring large amounts of data on the internet. We will use their GitHub accounts to document their growth and progress over time, ensuring they have a work portfolio demonstrating their skills upon completing the course. Python is recommended as a beginner-friendly programming language due to its intuitive syntax, availability of resources and libraries, cross-platform compatibility, and numerous use cases. The ChatGPT Python API and GitHub can promote critical thinking, collaboration, creativity, and communication skills in a more learner-centered approach.

### 4.2 Students Learning Performance and Evaluation

All student assignments and projects must be submitted to the GitHub version control platform, with five assignments and one mini-hackathon project to build the habit of turning ideas into work. Through the automatic version control tracking mechanism on GitHub, the submission tracking.

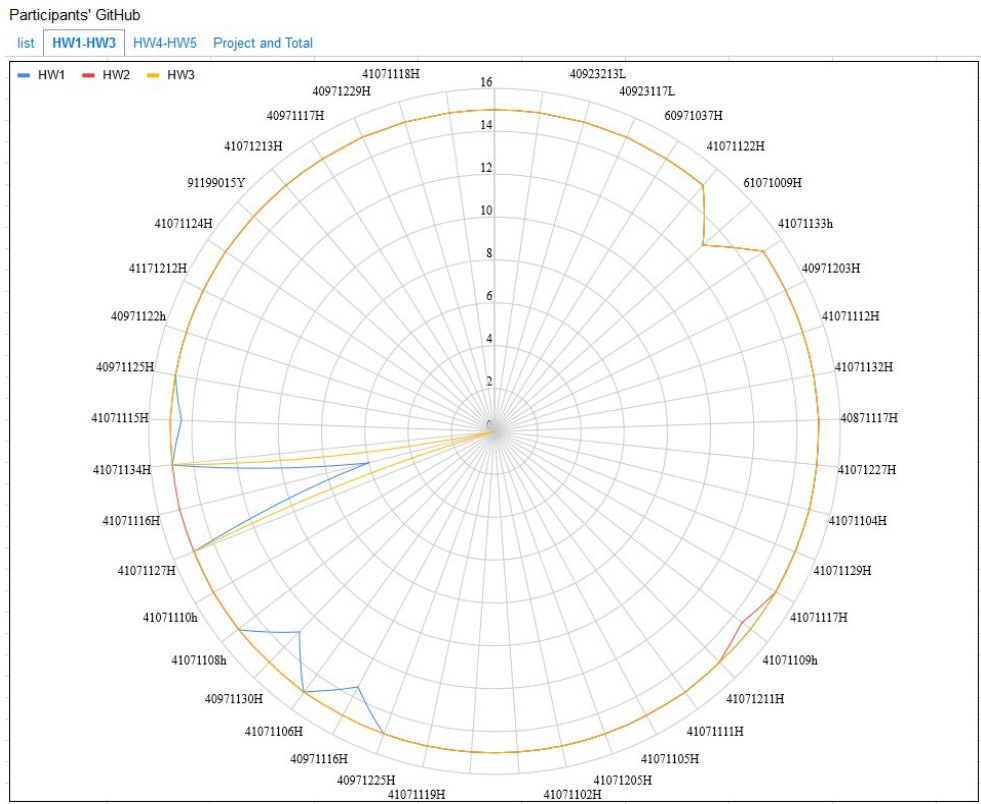

**Fig. 4.** The performance of all participants in all assignments.

1. Assignment 1, worth 15 points, is designed to use Sets to apply intersection union difference sets, allowing students to choose the dataset problem they want to solve for the first assignment. With the ChatGPT Python API, students can use natural language processing to analyze and understand their chosen dataset and generate insights and recommendations based on the data.
2. Assignment 2 is worth 15 points and is designed to confirm that all students are proficient in using JSON and Python Dict features with semi-structured data to solve problem styles. Analyze a mix of structured and semi-structured data. Verify that students can apply all syntax flexibly to the problem they are trying to solve. With the ChatGPT Python API, students can generate natural language descriptions of their solutions and analyze large amounts of text data to gain insights into the structure and organization of the data.
3. Assignment 3 is worth 15 points. It is designed to confirm that all students understand the inductive logic of text structure and can quickly batch-process large amounts of repetitive key content extraction through data regularization. Assure that all students can successfully use web crawling skills to extract large amounts of data of interest for analysis and application in their projects. With the ChatGPT Python API, students can use natural language processing to remove key content and insights from large amounts of text data, allowing them to quickly and efficiently analyze and organize the data.
4. Assignment 4, with a score of 15, is designed to confirm that all students can use data visualization and related analysis tools to visually present large amounts of data of interest to them and perform in-depth interpretation and analysis. With the ChatGPT Python API, students can generate natural language descriptions of their visualizations and gain insights into the patterns and relationships in the data.
5. Assignment 5, worth 15 points, is designed to confirm that all students can integrate the programmatic skills built in the previous four assignments to present a large amount of data on the Internet that they are interested in, along with textual exploration skills, and conduct an in-depth exploration. The final project was designed to confirm that all students were able to integrate the acquired skills from the previous ten weeks and that they were able to take a global view and explore the text of a large amount of data on the Internet that they were interested in, along with co-occurring network analysis skills. With the ChatGPT Python API, students can generate natural language descriptions of their findings and insights and collaborate with others on their analysis and exploration of the data.

6. The final project is worth 100 points and accounts for 25% of the total grade. The design is to design a user experience solution that incorporates all the acquired development skills into the problem they want to solve and the object they want to serve and visually represent the flow of use. With the ChatGPT Python API, students can generate natural language descriptions of their user experience solution and gain insights into the effectiveness of their solution based on user feedback and analysis of user behavior.

We can use the ChatGPT Python API in assignments and projects to provide students with personalized and adaptive learning experiences. The API can generate natural language responses, analyze text data, and provide targeted feedback and support. Additionally, it can develop customized learning materials for students based on their learning styles. Incorporating the ChatGPT API promotes a learner-centered approach to learning and helps students build their problem-solving skills.

|  | Very Much in Line | Still meets | Disagree |
|---|---|---|---|
| The course syllabus is arranged appropriately | 80% | 12.50% | 7.50% |
| Teaching to stimulate interest in learning | 75% | 15% | 10% |
| Teachers teach from the heart | 82.50% | 10% | 7.50% |
| Good interaction between teachers and students | 87.50% | 12.50% | 0% |
| The evaluation method is reasonable | 77.50% | 17.50% | 5% |

**Table 1.** Course Evaluation. There are forty-one students participated in the classroom feedback.

Table 1 shows that forty-one students participated in the classroom feedback, which showed that students thought that this teaching method could stimulate learning interest and allow good interaction between teachers and students in the classroom. Students can track each other's changes and progress after submitting their codes and get the most realistic results of learning effectiveness. The live interactive video recording in the classroom will analyze the actual process of students working on the tasks and let students tell their progress and changes. With the confirmation of Tinker Learning teaching mode, students can construct their knowledge body step by step with the assistance of Live Coding & Live Debug.

## 5 Discussion

This section will discuss several key ideas related to the teaching approach and learning process, such as Tinker Learning, Learner-Centered Instruction, Cone of Learning, and Creative Learning Spiral. These concepts emphasize hands-on, experiential learning and encourage students to explore and discover concepts independently. By implementing various learning methods, such as collaborative learning, hands-on projects, mini-hackathon, flipped teaching, live video recordings of teaching operations, and program examples, students can demonstrate their implementation strategies for applying technology to educational training.

The ChatGPT Python API has significant potential to enhance the learning experience for students and promote learner-centered instruction. By leveraging natural language processing, machine learning, and text analysis, educators can provide students with a more personalized and adaptive learning experience. The ChatGPT API can generate natural language responses, allowing more engaging and interactive interactions between students and the program. This can be especially beneficial for students who may require additional approval or who learn at a different pace from their peers. Additionally, the ChatGPT API can analyze large amounts of text data, providing educators with insights into students' understanding of the subject matter and identifying areas where students may struggle.

Using ChatGPT, educators can provide personalized support and guidance, generate personalized learning materials, and offer targeted feedback to help students improve. Overall, the ChatGPT Python API offers a powerful tool for educators to promote learner-centered instruction and provide students with a more personalized and adaptive learning experience, ultimately leading to more effective learning outcomes.

# 6   Conclusion

In our programming course, we have leveraged the ChatGPT Python API to enhance students' sense of accomplishment and promote deeper thinking through qualitative analysis. By introducing problem situations and demonstrating how to solve them using Python language and packages, students have gained hands-on experience and practical skills that they can use to solve similar problems. We have also implemented the Tinker Learning teaching strategy, which encourages students to actively participate in writing code and constructing their own knowledge body. Using the ChatGPT API, students have explored new avenues in text analysis and language processing, enabling them to analyze larger samples and gain deeper insights into the subject matter. This has led to a significant increase in students' sense of accomplishment and motivation to continue learning. By encouraging students to think about how to apply the tools demonstrated to solve real-life problems, we have promoted the development of the 4C skills: critical thinking, communication, collaboration, and creativity. Using the ChatGPT API, students have expanded their problem-solving skills and developed a deeper understanding of the subject matter. We can verify their growth and progress by analyzing their records on GitHub. Our teaching methodology has helped students develop the ability to "learn how to learn" and build their own knowledge body, leading to a more profound understanding of programming concepts and principles. Overall, the ChatGPT API has played a crucial role in enhancing our teaching strategy and promoting students' sense of accomplishment and motivation. By incorporating problem-solving strategies and encouraging active participation, we have developed students' 4C skills and equipped them with valuable skills that will serve them well in their future careers.